\documentclass[breaklinks]{aa} 
\usepackage{txfonts}
\usepackage{orcidlink}
\usepackage{bm}
\usepackage[font=small, skip=0pt]{caption}

\usepackage{graphicx}	
\usepackage{amsmath}	
\usepackage{amssymb}	
\usepackage{subfig}
\usepackage[font=footnotesize]{caption}
\usepackage{wrapfig}
\usepackage{gensymb}
\usepackage{multirow}
\usepackage{tabularx}
\usepackage{csquotes}

\usepackage{lscape}
\usepackage{stackengine}
\usepackage[dvipsnames]{xcolor}
\usepackage{tablefootnote}
\usepackage{enumitem}
\usepackage{longtable}
\usepackage{threeparttablex}
\usepackage{hhline}	
\usepackage{pdflscape}	
\usepackage{comment}
\usepackage{breqn}

\def\maxithirt{MAXI~J1348--630}

\def\grs{GRS~1915$+$105}

\usepackage{natbib,twoopt}
\usepackage[hyphenbreaks]{breakurl}

\DeclareMathAlphabet\mathzapf       {T1}{pzc} {mb} {it}
\bibpunct{(}{)}{;}{a}{}{,}             
\definecolor{cobalt}{rgb}{0.06, 0.2, 0.65}
\hypersetup{
  colorlinks,
  citecolor=cobalt,
  linkcolor=cobalt,
  urlcolor=cobalt,
}
\makeatletter
  \newcommandtwoopt{\citeads}[3][][]{\href{http://adsabs.harvard.edu/abs/#3}
    {\def\hyper@linkstart##1##2{}
     \let\hyper@linkend\@empty\citealp[#1][#2]{#3}}}
  \newcommandtwoopt{\citepads}[3][][]{\href{http://adsabs.harvard.edu/abs/#3}
    {\def\hyper@linkstart##1##2{}
     \let\hyper@linkend\@empty\citep[#1][#2]{#3}}}
  \newcommandtwoopt{\citetads}[3][][]{\href{http://adsabs.harvard.edu/abs/#3}
    {\def\hyper@linkstart##1##2{}
     \let\hyper@linkend\@empty\citet[#1][#2]{#3}}}
  \newcommandtwoopt{\citeyearads}[3][][]
    {\href{http://adsabs.harvard.edu/abs/#3}
    {\def\hyper@linkstart##1##2{}
     \let\hyper@linkend\@empty\citeyear[#1][#2]{#3}}}
\makeatother

\newcommand{\be}{\begin{equation}}
\newcommand{\en}{\end{equation}}

\def\ltsima{$\; \buildrel < \over \sim \;$}
\def\lsim{\lower.5ex\hbox{\ltsima}}
\def\gtsima{$\; \buildrel $\geq$ \over \sim \;$}
\def\gsim{\lower.5ex\hbox{\gtsima}}

\renewcommand{\linenumbers}{}

\begin{document} 

   \title{Rapid jet production and suppression during fast state transitions in the black hole X-ray binary \maxithirt{}}

   \authorrunning{Carotenuto et al.}
   \titlerunning{X-ray variability and transient jets in the black hole X-ray binary \maxithirt{}}

   \author{F.~Carotenuto\inst{1}\orcidlink{0000-0002-0426-3276} 
          \and
    L. Zhang\inst{2,3}\orcidlink{0000-0003-4498-9925}
          \and
          D. Altamirano\inst{3}\orcidlink{0000-0002-3422-0074} 
          \and
          P. Casella\inst{1}\orcidlink{0000-0002-0752-3301}
          \and
          S. Corbel\inst{4}\orcidlink{0000-0001-5538-5831}
          \and 
          J.C.A. Miller-Jones\inst{5}\orcidlink{0000-0003-3124-2814}}
   \institute{INAF Osservatorio Astronomico di Roma, Via Frascati 33, I-00078, Monte Porzio Catone (RM), Italy\\
              \email{francesco.carotenuto@inaf.it}
       \and
       Key Laboratory for Particle Astrophysics, Institute of High Energy Physics, Chinese Academy of Sciences, 19B Yuquan Road, Beijing 100049, People’s Republic of China
       \and
            School of Physics \& Astronomy, University of Southampton, SO17 1BJ, UK
            \and
            Université Paris Cité and Université Paris Saclay, CEA, CNRS, AIM, F-91190 Gif-sur-Yvette, France
            \and
            International Centre for Radio Astronomy Research -- Curtin University, GPO Box U1987, Perth, WA 6845, Australia}

   \date{Received XX; accepted XX}

\abstract{
Black hole X-ray binaries (BH XRBs) launch powerful relativistic jets during bright outburst phases. The properties of these outflows change dramatically between different spectral/accretion states. Compact jets are observed during the hard state and are quenched during the soft state, while discrete ejecta are mainly launched during the hard-to-soft state transition. Currently, we do not understand what triggers the formation/destruction of compact jets or the launch of discrete ejecta. In this context, finding a unique link between the jet evolution and the properties of the X-ray emission, such as its fast variability, would imply major progress in our understanding of the fundamental mechanisms that drive relativistic outflows in BH XRBs. Here we show that a brief, strong radio re-brightening during a predominantly soft state of the BH XRB \maxithirt{}  was contemporaneous with a significant increase in the X-ray rms variability observed with NICER in 2019. During this phase, the variability displayed significant changes and, at the same time, \maxithirt{} launched two relativistic discrete ejecta that we detected with the MeerKAT and ATCA radio-interferometers. We propose that short-lived compact jets were reactivated during this excursion to the hard-intermediate state and were switched off before the ejecta launch, a behavior that has been very rarely observed in these systems. Interestingly, with the caveat of gaps in our radio and X-ray coverage, we suggest a tentative correspondence between the launch of ejecta and the drop in X-ray rms variability in this source, while other typical X-ray signatures associated with discrete ejections are not detected. We discuss how these results provide us with insights into the complex dynamic coupling between the jets and hot corona in BH XRBs.}


   \maketitle

\section{Introduction}
\label{sec:intro}

Black hole X-ray binaries (BH XRBs) are systems in which a stellar-mass black hole accretes matter from a companion star through an accretion disk, and in which matter can be ejected in the form of relativistic jets. Due to their proximity, evolution timescales and wide range of accretion states, BH XRBs allow us to deeply explore the fundamental connection between the jets and the accretion process. During bright outburst phases, BH XRBs undergo spectral and timing state transitions that reflect changes in the structure and dynamics of the accretion flow. Following the current classification, four primary accretion states are identified \citep{Homan_belloni}: the hard state (HS), hard-intermediate state (HIMS), soft-intermediate state (SIMS), and soft state (SS). During the rise phase of an outburst, BH XRBs typically evolve from the HS through the HIMS and SIMS before reaching the SS. The decline phase generally follows the reverse sequence, with the system returning to the HS before going back to quiescence (e.g.\ \citealt{Remillard_xrb, Dunn_2010}). After decades of observations, there is now a well-established phenomenological connection between the accretion flow and jets in these systems. Different types of jets are produced in different accretion states \citep{Corbel_2004, Fender_belloni_gallo}. Compact jets, evolving with the accretion rate, emit self-absorbed synchrotron radiation from the radio through near-infrared and are observed during the HS \citep{Corbel_2000, Fender_2001, Markoff_2001, Corbel2002, Russell_2013b}. No jet activity is detected in the SS, as compact jets are strongly quenched before the state transition \citep{Fender_1999_quenching}. The quenching happens on dynamical (hours to days, \citealt{Russell_2020_break_frequency}) timescales and the emission decays first in the NIR, and subsequently in the radio. Conversely, discrete jet ejecta are launched during transitions between the HS and the SS, producing bright radio flares. During these events, the synchrotron emission usually evolves from initially self-absorbed to optically thin at radio frequencies (e.g.\ \citealt{Tetarenko2017}). The ejecta typically appear as bipolar plasma blobs moving outward from the core, often displaying apparently superluminal speeds \citep{Mirabel1994, Fender1999, Bright}. These components can propagate up to parsec scales far from the core, re-brightening and decelerating as they deposit energy in the nearby interstellar medium \citep{Corbel2002_xte, Corbel2005_h17, Espinasse_xray, Carotenuto_2022, Bahramian_2023}, with a long-term impact that has been characterized in a number of sources (e.g.\ \citealt{Motta_2025, BoschCabot2026}). Despite extensive studies (e.g.\ \citealt{Miller-Jones_h1743, Homan_qpo, Wood_2021}), a precise temporal association between changes in the inner accretion flow and ignition/suppression of compact jets, as well as the production of discrete ejecta, remains unclear. In particular, the uncertainty on the formation and launch of discrete ejecta largely stems from the absence of accurate measurements of their ejection dates and from the lack of sufficiently dense X-ray monitoring.

\maxithirt{} is a BH XRB discovered in January 2019 \citep{Yatabe2019}. During its 2019/2020 outburst, the source first completed a whole cycle in the Hardness Intensity Diagram (HID), with strong radio flaring at the first state transition \citep{Tominaga_1348, Carotenuto_atel}, and then displayed a sequence of hard state re-brightenings that lasted until September 2020. \maxithirt{} is located between $\sim$2.2 and 3.4 kpc \citep{Chauhan2021, Lamer_2021}. The orbital period of the system and the BH mass are currently unknown. 
In this paper, we combine previously published radio and X-ray timing data of the BH XRB \maxithirt{} to analyze in detail a specific short phase of the outburst to probe its complex accretion/ejection coupling during state transitions. The outburst phase that is the subject of this study was characterized by a brief excursion of the system from the soft state to the HIMS, with a fast switching on and off of compact jets, as well as the launch of discrete ejecta. The paper is structured as follows: the data collection is presented in Section \ref{sec:Data_collection}, while the comparison between the radio data and the X-ray timing results is presented in Section \ref{sec:results}. Finally, the main discussion and conclusions are presented in Sections \ref{sec:Discussion} and \ref{sec:Conclusions}.

\section{Data collection}
\label{sec:Data_collection}

In this work, we use the full radio and X-ray monitoring of the source during its 2019/2020 outburst \citep{Carotenuto2021}, including data from the MeerKAT radio-interferometer \citep{Jonas2016, Camilo2018}, the Australia Telescope Compact Array (ATCA, \citealt{Frater_1992}) and the Neutron Star Interior Composition Explorer (NICER, \citealt{Gendreau_2016}) on board the ISS. In more detail, \maxithirt{} was observed with MeerKAT at 1.28 GHz as part of the ThunderKAT Large Survey Programme \citep{ThunderKAT} with an approximately weekly cadence, collecting 48 15-min epochs between January 2019 and March 2020. \maxithirt{} was also monitored with ATCA for 31 epochs in total, from 2019 January to December (project codes C1199 and CX423). The array spanned multiple configurations during this time, but for each epoch data were recorded simultaneously at central frequencies of 5.5 GHz and
9.0 GHz, with 2 GHz of bandwidth at each frequency. The radio data were reduced using standard practices within the Common Astronomy Software Applications (CASA, \citealt{CASA_2022}), including flagging, calibration and imaging. When imaging ATCA data, we used a Briggs robust parameter of 0 to balance sensitivity and resolution \citep{Briggs_1995}, while we used a uniform weighting scheme to maximize the angular resolution of the MeerKAT images \citep{Carotenuto2021}.

On the other hand, NICER has extensively monitored the outburst of \maxithirt{} with almost daily cadence between 2019 January 26 and 2019 October 8, for a total of $\sim$ 274 ks. All observations were reprocessed using the NICER software tools {\tt nicerdas} version 6.0, distributed with {\tt heasoft} version 6.26. The data-reduction procedure was described in detail in \cite{Zhang2020}. While the full X-ray spectral-timing analysis has been presented in a number of papers \citep{Zhang2020, Belloni_2021, Garcia_2021, Zhang_2021, Zhang_2022, Alabarta_2025}, here we combine the X-ray timing information with the radio monitoring to build one of the clearest views of the interplay between jets and inflow during less-explored outburst phases.

\section{Results}
\label{sec:results}

\subsection{The outburst evolution}
\label{sec:The outburst evolution}

\begin{figure*}
  \center
  \includegraphics[width=\textwidth]{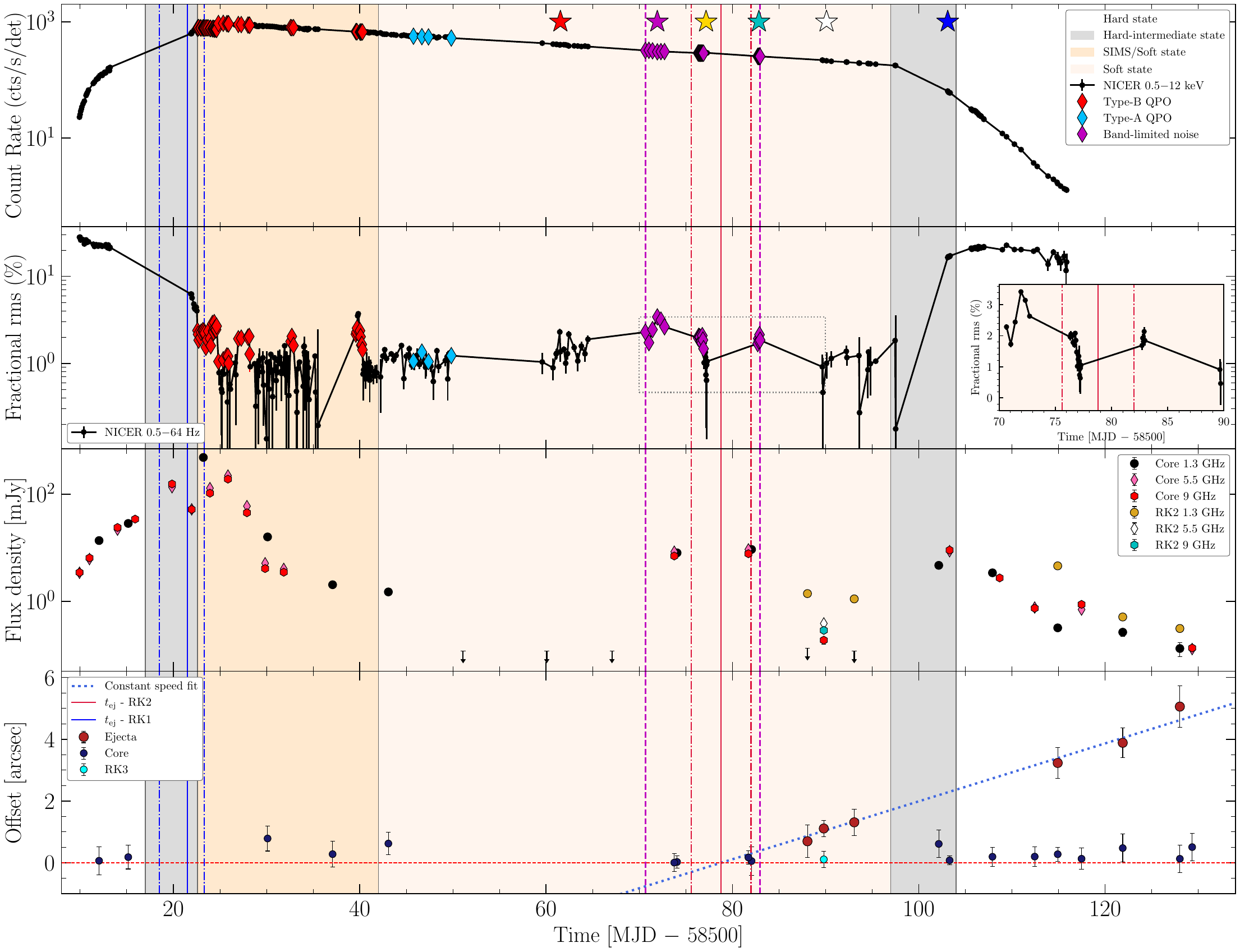}
  \captionsetup{font=small, skip=0pt}
  \caption{\emph{Top panel}: NICER 0.5--12 keV X-ray light curve (count rate) of \maxithirt{} during its 2019/2020 outburst. The white, light-gray, light-orange and light-pink regions denote, respectively the hard, HIMS, SIMS and soft state, from \protect\cite{Zhang2020}. The purple vertical dashed lines region marks the presence of intermittent strong band-limited noise during the soft state. Type-B and A QPOs are marked with red and light-blue points, respectively. The colored stars mark the times in the light curve at which the power spectra shown in Figure \ref{fig:PSD} were obtained. \emph{Second panel}: X-ray fractional rms variability, calculated in the 0.5–64 Hz frequency range. The inset on the right shows in detail the evolution of the variability between MJD 58570 and 58590.
  \emph{Third panel}: MeerKAT (1.3 GHz) and ATCA (5.5 and 9 GHz) core and RK2 radio light curves, from \protect\cite{Carotenuto2021}. \emph{Bottom panel}: Angular separation in arcsec between RK2 and \maxithirt{} from the same radio observations. Detections of the core are shown as dark blue points. The single detection of RK3 is also marked with a cyan point. A linear motion was used to fit the RK2 data. The red vertical line marks the RK2 inferred ejection date: $t_{\rm ej} = \rm MJD \ 58578.8 \pm 3.2$, where the two red dashed lines represent the upper and lower bounds of the $t_{\rm ej}$ confidence interval. The same is shown in blue for RK1, with $t_{\rm ej} = \rm MJD \ 58521.5^{+1.8}_{-3.0}$  from \protect\cite{Carotenuto_2022}.}
  \label{fig:lcurves}
\end{figure*}

\maxithirt{} displayed a rather classical BH XRB outburst, which has been thoroughly followed by dense X-ray and radio monitoring campaigns, as can be seen from Figure \ref{fig:lcurves}. In the top panel, we show the NICER (0.5--12 keV) count-rate light curve, with the spectral states highlighted, taken from \cite{Zhang2020}. \maxithirt{} was detected on MJD 58509 in the HS, and transitioned first to the HIMS on MJD 58517 while rising in flux, and then to the SIMS on MJD 58522.6, while completing the transition to the SS on MJD 58542 \citep{Zhang2020}. Subsequently, the system displayed a smooth decay in X-ray flux and transitioned back to the HIMS on MJD 58597, and back to the HS on MJD 58604 (see Figure \ref{fig:lcurves}). 

\begin{figure*}
  \centering
  \includegraphics[width=\textwidth]{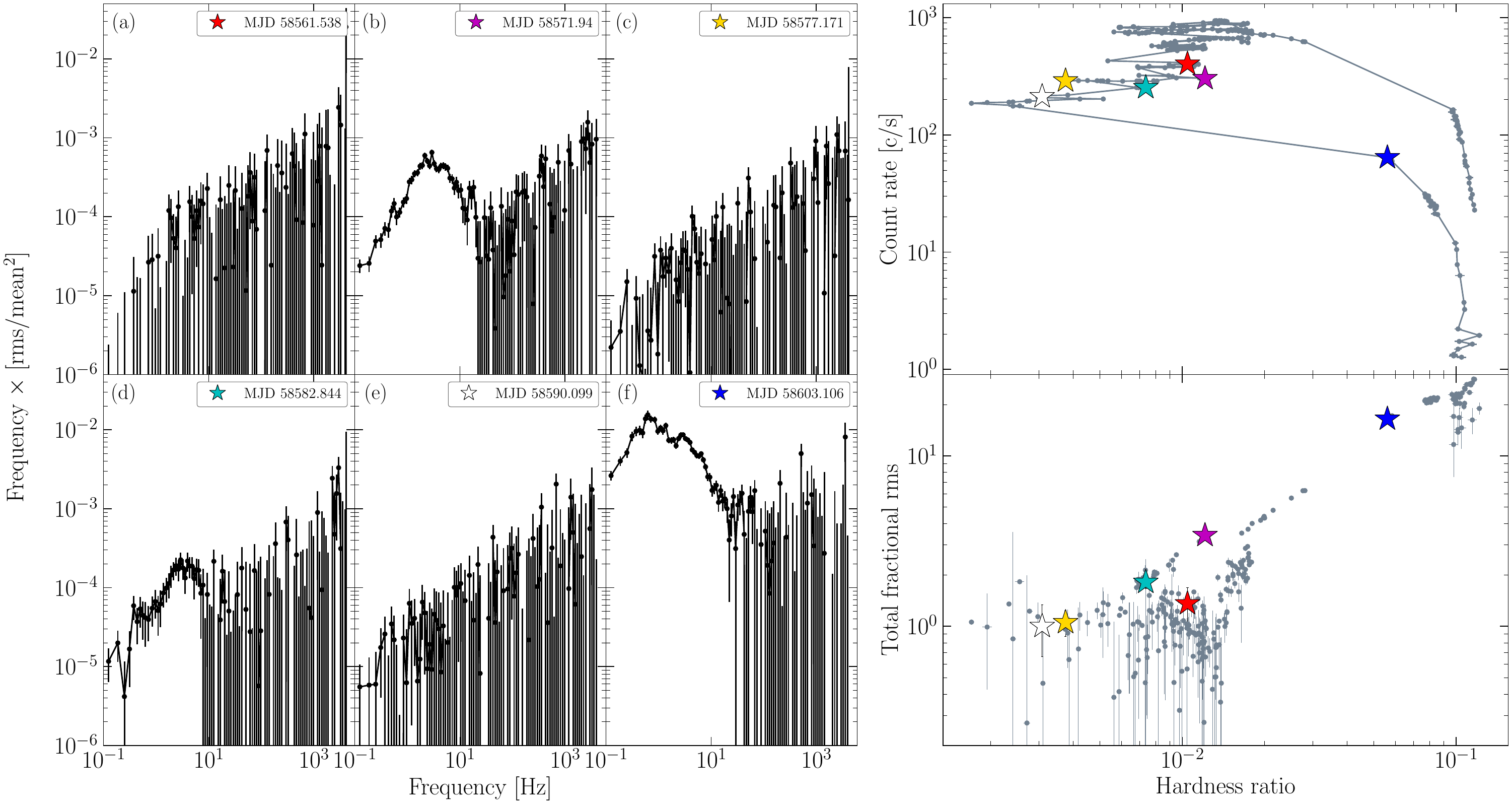}
  \captionsetup{font=small, skip=0pt}
  \caption{\emph{Left panels}: Representative NICER power spectra at different phases of the outburst, calculated in the 0.5--12 keV band and labeled from \textit{a} to \textit{f}. \textit{Right panels}: NICER hardness-intensity diagram (HID, top) and hardness-rms diagram (HRD, bottom) of the outburst. The hardness ratio is calculated as $(6 - 12)/(2 - 3.5)$ keV, while the total fractional rms is calculated in the 0.5–64 Hz frequency range (and 0.5--12 keV energy range). In both rows, the colored stars mark the position in the diagrams at which the power spectra were obtained. The presence of strong band-limited noise typical of the hard state supports the reactivation of compact jets during the short HIMS phase around MJD 58570. The two drops in X-ray rms variability happening after 58576 and after MJD 58582 could be consistent with the launch of the RK2 and RK3 discrete ejecta, respectively.}
  \label{fig:PSD}
\end{figure*}

For all the NICER epochs, we also consider the evolution of the 0.5--64 Hz fractional rms variability (within the whole NICER 0.5--12 keV energy range), which is shown in the second panel from the top in Figure \ref{fig:lcurves} \citep{Zhang2020}. As expected, higher variability (up to $\sim$30\%) was observed in the HS, while the variability is much less in the IMS and SS ($\lesssim$5\%). Different types of low-frequency Quasi Periodic Oscillations (QPOs) have been detected during the outburst. Type-A and -B QPOs are marked in Figure \ref{fig:lcurves}, while faint Type-C QPOs were sometimes detected in the beginning (MJD 58509-58522.4) and end (MJD 58603-58615) of the main outburst, along with band-limited noise \citep{Zhang2020, Zhang_2021}. The outburst evolution in the X-rays can also be followed on the NICER Hardness-Intensity Diagram (HID) displayed in the top right panel of Figure  \ref{fig:PSD}. The same evolution is displayed on the Hardness-rms diagram (HRD) in the bottom right panel of Figure \ref{fig:PSD}. 

Coming back to Figure \ref{fig:lcurves}, we show the radio light curve at the position of \maxithirt{} of the same period in the third panel from the top. Our multi-frequency (1.3, 5.5 and 9 GHz) radio observations traced the evolution of compact jets, capturing their initial brightening during the HS, their quenching as the system entered the SS, and subsequent re-emergence during HS re-flares \citep{Carotenuto2021}. Two subsequent, single-sided, approaching discrete ejecta have been detected and tracked as they moved away from the compact object in the same direction \citep{Carotenuto2021, Carotenuto_2022}, displaying the third and fourth highest proper motion ($\gtrsim$100 mas d$^{-1}$) ever observed so far among BH XRBs (after 4U~1543--47, \citealt{Zhang_2025}). Since it is highly relevant for this work, the angular separation of the second ejecta (labeled Radio Knot 2, RK2) is shown in the fourth panel of Figure \ref{fig:lcurves}. While we expect each approaching ejection component to have a corresponding receding component, no receding ejecta from \maxithirt{} have been detected so far.

\begin{figure}
\begin{center}
\includegraphics[width=\columnwidth]{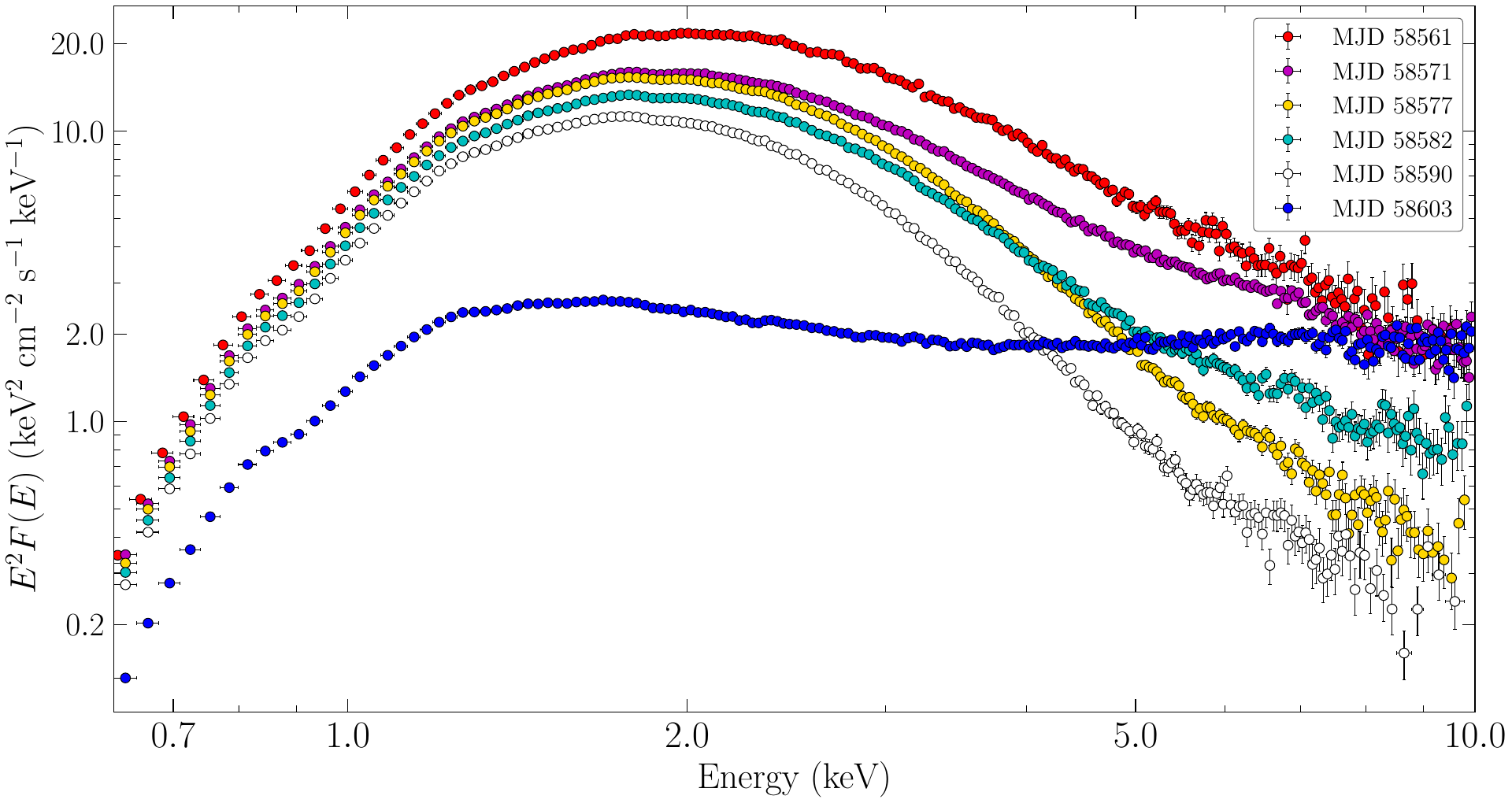}
\caption{NICER 0.6--10 keV unfolded energy spectra associated to the PSDs shown in Figure \ref{fig:PSD}, with the same color-code. The unfolded spectra, dominated by the disk component, have been deconvolved against a power law with $\Gamma = 2$, without any fitting, and are shown for illustrative purposes. We note that the spectra with broadband noise (magenta and cyan points) have slightly harder tails at high energies than spectra without variability (e.g. yellow and white points), supporting the scenario in which the variability is connected to the hot corona, which may disappear in the aftermath of a discrete ejection.}
\label{fig:spectra}
\end{center}
\end{figure}

\subsection{The X-ray properties between MJD 58570 and 58590}
\label{sec:The X-ray properties between MJD 58570 and 58590}

In this work, we only consider the outburst phase between MJD 58570 and 58590, which starts in the SS and then displays a complex evolution, including multiple possible state transitions not yet explored in detail. This behavior is not apparent from the smoothly-decaying X-ray light curve alone, while the HID (in Figure \ref{fig:PSD}) shows very fast shifts in the hardness ratio at relatively constant count rates. From a timing perspective, the X-ray variability displays strikingly different properties during this phase \citep{Zhang2020}. In the main left panel of Figure \ref{fig:PSD}, we show six Poisson-subtracted power spectra linked, through colored stars, to different points in the light curve, and labeled with letters from \textit{a} to \textit{f}. The distributions of the power spectral densities (PSD) are representative of the evolution of the source in this phase, as specified in the following, and there is little evolution between PSDs within the period indicated with each star. While, as mentioned before, there is low variability in the SS (panel \textit{a}), significant band-limited noise ($\sim$3\% rms variability) is constantly observed between MJD 58570 and 58576 (panel \textit{b}, and the region enclosed by the purple vertical dashed lines in Fig.\ \ref{fig:PSD}), sometimes with an intermittent weak, unclassified QPO around 18 Hz \citep{Zhang2020}, similar to the PSD that are typically observed in the HIMS (e.g.\ \citealt{Homan_belloni}). This variability is then absent in the following days (panel \textit{c}), until MJD 58582, when it briefly re-appears before being again undetected until MJD 58603 (panels \textit{d} and \textit{e}), when \maxithirt{} re-enters the HS (panel \textit{f}). It is worth mentioning that, as can be seen from the HRD in Figure \ref{fig:PSD}, the appearance of the variability places the source on the upper branch of the diagram, which is typically tracked by sources in the HS and HIMS (the lower branch is tracked in the SIMS, e.g.\ \citealt{Belloni_2010}). Throughout this particular outburst phase, \maxithirt{} oscillates between the upper and the lower branch of the HRD.

In order to check for potential spectral differences between epochs with different levels of variability, we extracted the energy spectra associated to the six panels of Figure \ref{fig:PSD}, and we plot them in Figure \ref{fig:spectra}, using the same color code between the plots. The unfolded spectra, rebinned to have at least 30 counts per bin, have been deconvolved using XSPEC \citep{Arnaud_xspec} against a reference power-law model with photon index $\Gamma = 2$ and are shown for illustrative purposes, without any detailed spectral fitting (see \citealt{Zhang2020}). This representation ($E^2 F(E)$, in units of keV$^2$ (photons cm$^{-2}$ s$^{-1}$ keV$^{-1}$) is used to facilitate comparison of the spectral shapes between epochs.
The spectra are mostly dominated by the disk component, except for the last observation of MJD 58603, which flattens to a simple HS $\Gamma = 2$ spectrum. 
From a visual inspection, we note that the spectra associated with broadband noise (magenta and cyan points) exhibit a slightly harder high-energy tail compared to those observed during epochs of low or absent variability (e.g. the yellow and white points).

\subsection{Evidence for additional discrete ejecta from \maxithirt{}}
\label{sec:Evidence for additional discrete ejecta}

\begin{figure}
\begin{center}
\includegraphics[width=\columnwidth]{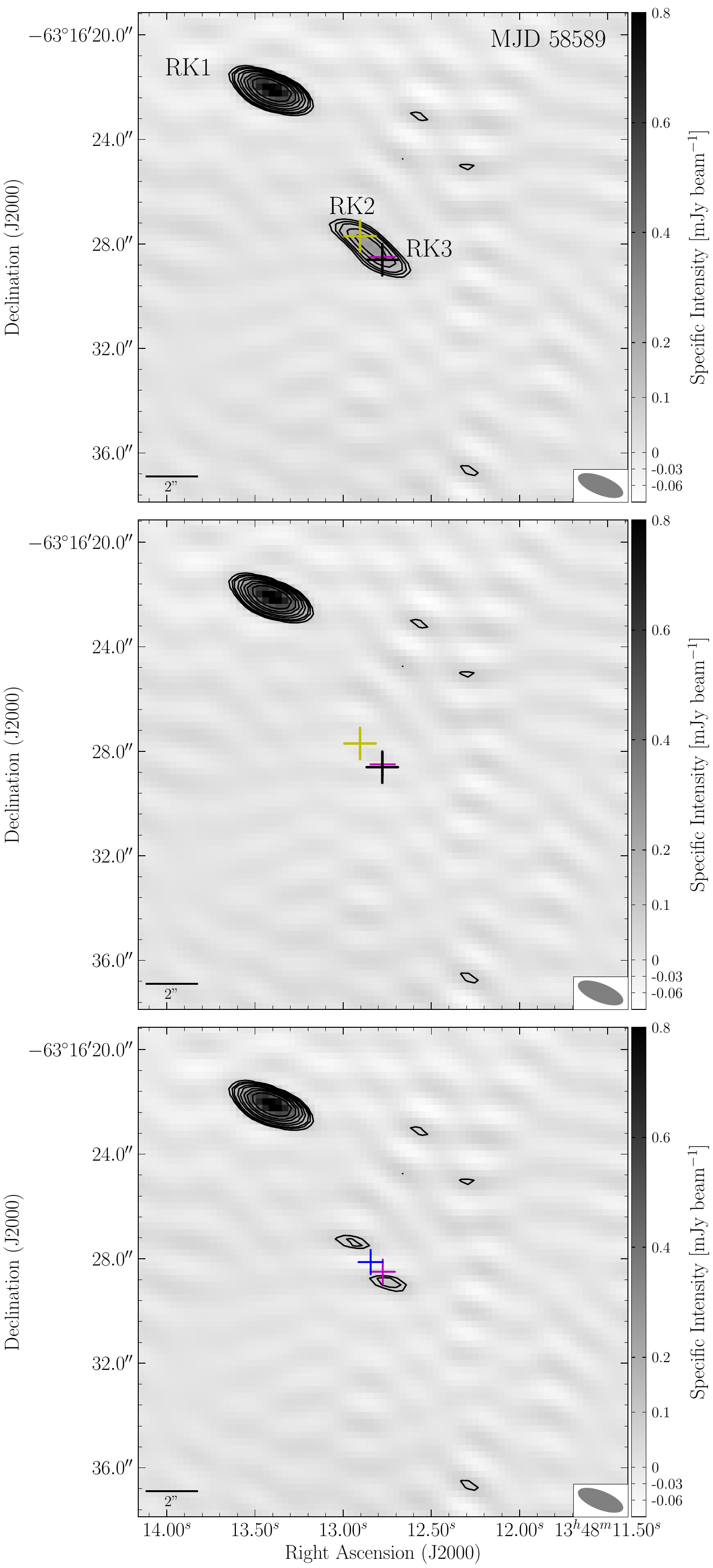}
\caption{ATCA 9 GHz image of the 2019 April 16 observation (MJD 58589), with two radio sources identified as RK1 and RK2 in \cite{Carotenuto2021}. Contours start at 3 times the rms ($\sim$25 $\mu$Jy beam$^{-1}$). The position of \maxithirt{} is marked with a magenta cross. \textit{Top:} the original RK2 emission is produced by two point-sources whose fitted positions are marked with a yellow and black cross. We identify the yellow cross with the true position of RK2, at an angular distance of $\sim$1.1 arcsec from the core, while the black cross is an additional component (RK3) or core emission from \maxithirt{}. \textit{Center:} residual image after the image-plane subtraction of the two above-mentioned point-sources, showing only gaussian-like noise at the position of \maxithirt{}. \textit{Bottom:} residual image after the image-plane subtraction of a single point-source with the location free to vary, and marked with a blue cross, showing residuals above three sigma level near \maxithirt{}.}
\label{fig:atca_images}
\end{center}
\end{figure}

From the radio point of view, strong radio emission originating from the core was detected with MeerKAT and ATCA on both MJD 58573 and 58582, reaching nearly 10 mJy at 1.3 GHz on MJD 58582. For both the ATCA detections, the radio spectral index, defined as $\alpha$, where the radio flux density follows $S_{\nu} \propto \nu^{\alpha}$, was $-0.37 \pm 0.04$. The MeerKAT flux density at 1.3 GHz was intermediate between the simultaneous ATCA 5.5 and 9 GHz detections, likely hinting at a spectrum that was not fully optically thin. The emission subsequently declined below the detection threshold in the following observation six days later \citep{Carotenuto2021}, quenched at least by a factor 75. Interestingly, \maxithirt{} launched its second discrete ejection close to the end of this outburst period. The approaching component, labeled RK2 in \cite{Carotenuto2021}, was spatially resolved with MeerKAT and ATCA, and was followed up in its motion up to an angular distance of $\sim$5 arcsec from the core. 

However, in this work, we re-consider the 9 GHz ATCA observation taken on MJD 58589 (2019 April 16), which is our highest angular resolution observation of RK2. This re-inspection reveals that the source originally identified as RK2 is instead a resolved radio source that can be well-described by the partial overlap of two point-like sources. The re-inspection of the data, which were thought to only contain evidence of RK1 and RK2, has been primarily motivated by a re-consideration of all radio data taken during this period characterized by fast state transitions, which are crucial to infer the ejecta launch date and place them in context with the X-ray timing results. Here we present a revised simultaneous fitting of the two point sources performed with the CASA {\tt imfit} task. The positions of the two point sources are shown in Figure \ref{fig:atca_images} with a yellow and black cross. The fit residuals are shown on the center panel of \ref{fig:atca_images}. We also present a fit with a single point source, with the position left free to vary, and marked as a blue cross in the bottom panel of Figure \ref{fig:atca_images}. This image shows significant residuals around the inferred location of the single component, implying that the observed emission cannot be produced by a single un-resolved source.
 
Considering the top panel of Figure \ref{fig:atca_images}, the first source (yellow cross) has a flux density of $0.24 \pm 0.01$ mJy beam$^{-1}$, and we identify it as the true emission from RK2, now at a revised angular distance of $1.1 \pm 0.3$ arcsec from the core. The second source (black cross) is located at a position consistent with \maxithirt{} ($0.1 \pm 0.3$ arcsec), with a flux density of $0.23 \pm 0.01$ mJy beam$^{-1}$. We propose that this source is an additional discrete ejection launched by \maxithirt{} after RK2, therefore no earlier than MJD 58578, and before MJD 58589, the day of the ATCA observation. We refer to it as RK3.
The larger ATCA beam ($2.5\arcsec \times 1\arcsec$) at lower frequencies makes it impossible to repeat the same fitting at lower frequencies. Therefore, to obtain spectral information on the two sources, we directly fitted the two components in the visibility plane. Specifically, we used {\tt uvmultifit}
\citep{uvmultifit} to simultaneously fit three point sources (including the RK1 component) in the visibility plane, considering both the 5.5 and 9 GHz ATCA sub-bands, and fixing the three components at the locations obtained from the image plane fitting at 9 GHz. For RK2, we obtain a flux density  $0.29 \pm 0.02$ mJy beam$^{-1}$ at 7.25 GHz, with a spectral index $\alpha = -0.6 \pm 0.2$ (where the radio flux density follows $S_{\nu} \propto \nu^{\alpha}$). For RK3, we obtain, instead, a flux density of $0.31 \pm 0.02$ mJy beam$^{-1}$ at 7.25 GHz, with a spectral index $\alpha = -0.8 \pm 0.2$. These results imply steep radio spectra both for RK2 and RK3, which are consistent with optically thin synchrotron radiation emitted by discrete jet ejecta from BH XRBs (e.g.\ \citealt{Fender2006}).

\section{Discussion}
\label{sec:Discussion}

The combination of the X-ray variability properties and the spectral hardening suggests that \maxithirt{} underwent a very short transition from the SS to the HIMS. We see no evidence of the source passing through the SIMS, as no Type-B QPOs have been detected in this phase. However, we cannot exclude that this may have happened during the X-ray data gaps or that the QPOs are present in the cross-spectrum as it has been seen in other sources \citep{Jin_2025}, but a similar type of analysis is beyond the scope of this paper). This behavior is not completely new among BH XRBs. Several systems have been observed to undergo rapid transitions between the SS and HIMS/SIMS (e.g.\ \citealt{Homan_2001, Fender_2009}), with multiple radio flares accompanying repeated crossings of the upper branch of the hardness–intensity diagram (e.g.\ \citealt{Brocksopp_2001, Tetarenko2017}). However, in the case of \maxithirt{} the presence of a PSD dominated by band-limited noise, combined with the detection of bright radio core emission, supports the hypothesis that compact jets were reactivated during the short HIMS phase, between MJD 58570 and 58576, before the detection of discrete ejecta. Remarkably, our data combined allow us to witness the full jet cycle typical of BH XRBs (and possibly also of other accreting compact objects, e.g.\ \citealt{Migliari_2006}), which includes the ignition and suppression of compact jets, and the formation and acceleration of discrete ejecta. We discuss these aspects in the following, starting with compact jets.

\subsection{Short-lived compact jets}
\label{sec:Short-lived compact jets}

The reactivation of compact jets during a short-lived HIMS in the middle of the SS is rarely observed in BH XRBs. It is unlikely that this phenomenology is unique to \maxithirt{}, while it is more likely that in this case the quality of the multi-wavelength  (radio and X-ray timing) information gives us unique access into the source's rapidly evolving accretion/ejection coupling. In principle, the strong radio emission could be attributed to the flare associated with the subsequent launch of the ejecta. However, the emission is detected for more than 8 days \citep{Carotenuto2021}, which is much larger than the typical $\sim$hours-days timescale of a flare \citep{Tetarenko2017, Bright, Fender_2023}. According to the standard model, compact jets emit self-absorbed synchrotron radiation \citep{Blandford_Konigl}. The typical compact jet spectrum is flat or slightly inverted at radio frequencies, with a turnover frequency to a fully optically thin spectrum observed in the near-IR band during the hard state \citep{Corbel2002, Russell_2013b}. During their formation, the compact jets initially turn on in the radio band with an optically thin spectrum, which later evolves into optically-thick synchrotron emission as the spectral break moves to higher frequencies with time \citep{Corbel2013_IR}. Interestingly, in the bright 2015 outburst of V404 Cyg (not a standard soft-to-hard transition), the optical variability was observed to significantly rise during the compact jet formation, as the optical/IR jet base re-brightened \citep{Gandhi_2017}.

In this context, the optically thin spectrum ($\alpha = -0.37 \pm 0.04$) could be consistent with the short duration of the HIMS, as compact jets were not active for enough time to build up a flat radio spectrum. This scenario is also broadly consistent with the presence of long-lasting X-ray variability (MJD 58570--58576) and slightly harder spectra than in the preceding and following epochs (Figures \ref{fig:PSD} and \ref{fig:spectra}). It is also possible that the optically thin MeerKAT/ATCA detections are due to the flares associated with the discrete ejections, which were not covered with a sufficient cadence. To our knowledge, the only other example of brief compact jet reactivation has been reported for the 2002 outburst of the BH XRB 4U~1543--47 \citep{Russell_2020_1543}. The authors report the detection of an IR flare as the source briefly returned to the SIMS from the SS. This flare suggested compact jets reactivated during this brief (5 days) return to the SIMS, before the source transitioned back to the SS. The IR re-brightening was contemporaneous with the appearance of strong Type-B QPOs, an increase in the rms variability and a slight spectral hardening \citep{Russell_2020_1543}. While this is similar to \maxithirt{}, there are clear differences: in our case, the source did not show evidence of passing through the SIMS, as Type-B QPOs were not detected in this phase; more importantly, \maxithirt{} launched discrete ejecta that were not detected in 4U~1543--47. However, we cannot completely rule out the launch of discrete ejecta in 2002, since, while 4U~1543--47 was monitored at radio wavelengths during the outburst, the ATCA radio upper limits of $<3$ mJy at 4.8 and 8.6 GHz are not particularly constraining. We also note that 4U~1543--47, which is at a larger distance than \maxithirt{}, did produce large-scale discrete ejecta in more recent outbursts, with flux densities below the mJy level \citep{Zhang_2025}.

\subsection{Discrete ejecta and X-ray variability}
\label{sec:Discrete ejecta and X-ray variability}

From the discussion in Sections \ref{sec:The X-ray properties between MJD 58570 and 58590} and \ref{sec:Evidence for additional discrete ejecta}, we have evidence that \maxithirt{} produced two jets during the time interval between MJD 58570 and 58590, while the system exhibited significant changes in the characteristics of the X-ray variability (see Section \ref{sec:The X-ray properties between MJD 58570 and 58590}). While RK2 is spatially resolved from the core in multiple observations, we only have a clear detection of the second jet in our 9 GHz ATCA data, on MJD 58589. The detected radio emission could originate from either a very short-lived compact jet or from a newly ejected component. From its optically thin spectrum and from the fact that it was detected during a period in which \maxithirt{} was extremely soft in the X-rays, we propose that this source is an additional discrete ejection launched by \maxithirt{}, and we refer to it as RK3.

If indeed a discrete ejection, this component had to be launched after the ejection of RK2 and before MJD 58589, the day of the ATCA observation. We have no direct evidence of RK3 in any other MeerKAT or ATCA observation after MJD 58589, even though we note that its emission, if present, could have been confused with RK2 in the MeerKAT epochs following MJD 58574 (due to the MeerKAT $\sim$$5^{\prime\prime}$ PSF, which is much larger than the $\sim$1\arcsec separation between the two ejecta). The four angular separation points after MJD 58590 could be from a single ejection (either RK2 or RK3), or from the two components unresolved. If we consider the ATCA detection of RK3, it appears unlikely that the following detections are all uniquely from RK3, since this would imply a strong deceleration followed by a linear motion. Alternatively, some of the points (one to three) could be from RK2 while the remaining from RK3, but we also deem this scenario as unlikely, given that it would require alternate detections of single components, which would have to re-brighten and fade at different times. Therefore, the most reasonable conclusion is that points at MJD $> 58590$ are dominated by RK2, and that its propagation could be described by a simple constant speed motion. Under such an assumption, we present revised fits for its proper motion, and we show the evolution of the angular separation of its approaching component in the bottom panel of Figure \ref{fig:lcurves}. A simple linear fit yields a very high proper motion of $\mu = 94 \pm 12$ mas day$^{-1}$, similar to the first ejecta \citep{Carotenuto2021}. This corresponds to a projected space velocity of $1.2 - 1.9 \ c$ for distances between 2.2 and 3.4 kpc, implying that RK2 is intrinsically relativistic. The inferred ejection date $t_{\rm ej} =$ MJD 58578.8 $\pm$ 3.2 points towards an ejection happening contemporaneously with the drop in fractional rms variability - caused by the disappearance of the band-limited noise - and the rapid return of the system to a pure SS. Since the $1\sigma$ 3.2 days uncertainty on $t_{\rm ej}$ is not negligible, it is less likely, but not completely impossible, that the launch of RK2 could be associated with the second drop in rms variability, which happened after MJD 58582.

In principle, given its detection at the core location, RK3 could be consistent with being the receding component of RK2. However, both RK2 and RK3 are detected at a very similar flux density level during the observation on MJD 58589. If they were the approaching/receding pair of the same ejection, we would expect different flux densities from the two components, which would be seen at different stages of their evolution due the difference in light-travel time. For a single observation, assuming that the plasmoids are freely (linearly) expanding, \cite{MJ_2004_formalism} derived a formalism to compute the expected flux density ratio: $S_{\rm app}/S_{\rm rec} =  [(1+\beta\cos\theta)/(1-\beta\cos\theta)]^{k-p}$, where $\theta$ is the inclination angle of the jet axis, $\beta$ the intrinsic speed in units of $c$, $k = 3$ for discrete ejecta, and $p = 1-2\alpha$ is the electron population index.. Assuming a range of possible parameters ($\theta = 30\sim50 $ deg, $\Gamma = 1.5\sim2.5$, and a spectral index of $-0.6$, e.g.\ \citealt{Carotenuto_2022}), the formalism yields flux density ratios in the range $2\sim8$, which are not observed in our ATCA data. A slower jet would relax the flux ratio constraint, but would not be consistent with the observations, which show that RK2 has a very similar proper motion to RK1. Therefore we deem more likely that RK3 is an additional approaching component rather than the receding component of RK2.

Given that RK3 was detected at the location of the core of \maxithirt{} on MJD 58589, and expecting a proper motion of the same order of magnitude as RK1 and RK2 (but not necessarily similar), we can tentatively place its launch date between MJD 58582 and 58589.
This opens up the possibility that, also in the case of RK3, its launch could be connected to the disappearance of rms variability between MJD 58582 and 58590, at the end of a very short re-hardening phase, when the system moved back to leftmost part of the HID (white star in Figure \ref{fig:PSD}). We note that there is a gap in the NICER data between MJD 58582 and 58590; therefore, the drop in variability could have happened anytime in this time period, which is broadly consistent with the estimated time range for the launch of RK3. During the same period, strong optically thin radio emission was also detected on MJD 58582, possibly consistent with a flare associated with its ejection. However, we caution the reader that, due its single detection at the core location, we cannot completely rule out RK3 being a quenching compact core jet and not a discrete ejection. 

Overall, we propose that the launch of the RK2 and RK3 discrete ejecta could be associated with the only X-ray signature that we observe during this phase, namely the change in PSD shape with the sudden drop of the level of X-ray rms variability (from 2$\sim$3\% to $<0.5$\%, see the inset in the second panel of Figure \ref{fig:lcurves}). 
Moreover, another variability drop is observed within two days of the inferred ejection date MJD $58521.5^{+1.8}_{-3.0}$ of the first ejection RK1 (see Figure \ref{fig:lcurves}, \citealt{Zhang_2021, Carotenuto_2022}). Therefore, in \maxithirt{}, we may have three instances where a drop in fractional variability occurs contemporaneously (within a few days) with the launching of discrete ejecta. This correspondence is inferred from three distinct, subsequent approaching jet components (caveat RK3 being an ejection and not a short-lived compact jet), in the absence of any other unambiguous signature in the X-ray emission, such as spectral changes and/or QPOs, suggesting that the rms drop itself may be the clearest available X-ray signature of the discrete ejection events. We must remark that the absence of other X-ray signatures is convincing, but not completely irrefutable due to the gaps in the NICER coverage.

While a link between drops in variability and jet ejections has been suggested in other sources (e.g.\ \citealt{Fender_2009, Miller-Jones_h1743}), \maxithirt{} offers one of the clearest views on this connection, which would be a manifestation of the strong coupling that exists between the hot corona and the jets in these sources. The hot corona is the Comptonization region responsible for the non-thermal X-ray emission observed in these systems \citep{Zdziarski_corona}, whose geometry is still the subject of debate, and which is thought to be deeply connected to jets in BH XRBs, both for the evolution of compact jets during the hard state and for its possible disappearance when ejecta are launched (e.g.\ \citealt{Rodriguez_2003, Markoff_corona, Homan_2013, Mendez_2022}). The band-limited noise is suggested to be produced from propagating fluctuations in mass accretion rate within a hot inner flow (e.g.\ \citealt{Ingram_2016}). In this context, if the corona is partially or totally expelled during the ejection, the simultaneous drop in rms variability may reflect the removal of coronal material near the compact object. This mechanism would thus provide a promising X-ray signature for identifying the launch of discrete ejecta (e.g.\ \citealt{Miller-Jones_h1743, Russell_2020_1543}). The loss of coronal material would also be consistent with the (marginal) spectral softening observed between epochs preceding and following a discrete ejection event, i.e.\ panels \textit{(b)}-\textit{(c)}, and panels \textit{(d)}-\textit{(e)} in Figure \ref{fig:PSD}., since the coronal emission dominates at higher energies (e.g. \citealt{Prat_2010}).

To date, multiple X-ray intensity, spectral, and timing signatures have been proposed as indicators of the launching of relativistic jets (e.g.\ \citealt{Belloni_Motta_2016}). Beyond the drop in rms variability, possible signatures include soft X-ray dips, as observed in \grs{} (e.g.\ \citealt{Vadawale_2003}); the appearance of specific features in the X-ray power spectrum, such as the Type-B QPOs (e.g.\ \citealt{Soleri2008, Fender_2009}); the transition from Type-C to Type-B QPOs (e.g.\ \citealt{Homan_qpo}); and changes in the sign of phase lags at the QPO frequency, along with simultaneous radio emission \citep{Mendez_2022}. These X-ray observables are linked to the properties of the corona, such as its geometry, size or magnetic field. Specifically, in many systems the ejections have been inferred to happen hours or even days before the detection of Type-B QPOs \citep{Miller-Jones_h1743, Russell_2019, Carotenuto_2024}, making the causality relation between these two phenomena particularly unclear. Type-B QPOs may be produced by the corona alone in the SIMS, but they may not directly trace the acceleration of relativistic plasmoids.

Changes in coronal geometry across spectral states have also been suggested. Based on the evolution of time lags, it has been argued that during the rising HS, the corona contracts and becomes less vertically extended \citep{Kara_2019}, while in the HIMS/SIMS, observations show that the coronal height increases, up to $\sim$100 $R_{\rm g}$, possibly representing coronal material being ejected \citep{Wang_2022, Liu_2022}, which is consistent with the detection of contemporaneous soft dips \citep{Vadawale_2003}. Consistent with this picture, the spectral modelling of NuSTAR data of \maxithirt{} on MJD 58577 (which lies within the inferred ejection date range for RK2) reveals an increased corona height (up to $\sim$5 $R_{\rm g}$ from an average of $\sim$2 $R_{\rm g}$)  \citep{Davidson_2025}. \cite{Mendez_2022} and \cite{Garcia_2022} proposed that, in \grs{}, the corona and the jets are, at different times, the same physical component, while outflowing coronas were also proposed to explain observed correlations between radio emission, time lags and X-ray spectral evolution (e.g.\ \citealt{Kylafis_2008, Reig_2018}). 

It is important to note that recent X-ray polarimetric results appear to contradict this scenario, showing that the coronal geometry appears unchanged throughout the HS and the full transition to the SS in Swift~J1727.8-1613, at least over timescales of months  \citep{Ingram_2024}. This implies that, if a change in coronal geometry really happens during each ejection, the HS geometry must be recovered each time on hours-to-days timescales to be consistent with the polarimetric findings. 
At the extreme, these systems can also produce multiple ejecta within hours/days \citep{Brocksopp, Miller-Jones2019, Wood_2021, Wood_2025}. Interestingly, recent GR particle-in-cell simulations suggested a scenario where radio ejections observed during state transitions are produced from the launching of magnetic loops along field lines coupling the accretion disk to the central BH in the inner corona \citep{Mehlhaff_2025}. In this picture, magnetic reconnection accelerates discrete ejecta, and each magnetic loop corresponds to a single ejection; thus multiple ejection events are possible \citep{Tagger_2004, Mehlhaff_2025}.

\section{Conclusions}

\label{sec:Conclusions}
Although the precise causal connection between the changes in the accretion inflow and the ignition/suppression of compact jets, as well as the launch of discrete ejecta, is still unclear, our results on \maxithirt{} represent one of the new insights on the precise sequence of timing signatures and jet activity in these sources. We propose that each discrete ejection could be associated with a drop in X-ray rms variability, and this phenomenology could be explained by the ejection of coronal material. Our findings further motivate the need for high-angular resolution radio observations coupled with a dense and uniform X-ray timing coverage of BH XRBs in outburst. Indeed, milli-arcsec resolution observations using very long baseline interferometry (VLBI) can spatially resolve multiple ejecta and obtain the most accurate information on the ejection dates \citep{Miller-Jones2019, Wood_2021, Wood_2023, Wood_2024, Wood_2025}, which can then be associated with changes in the spectral and timing properties of the X-ray emission in these systems. This is essential to make progress, as it will allow us to better understand the role of the various physical components of the system (e.g.\ disk, corona, jet base) in the production and acceleration of discrete ejecta, and to ultimately constrain the jet launching mechanism. 

\bigskip
\noindent
\textbf{Data Availability:} The radio-interferometric data used in this work have been presented in \cite{Carotenuto2021}. The NICER data have been published in \cite{Zhang2020} and are available in the HEASARC database.

\begin{acknowledgements}
We thank the anonymous referee for the feedback, which significantly improved the paper.
This project made use of \textsc{matplotlib} \citep{matplotlib}, \textsc{numpy} \citep{harris2020array} and Overleaf (\url{https://www.overleaf.com}). This research also made use of APLpy, an open-source plotting package for Python \citep{aplpy2012, aplpy2019}.

\end{acknowledgements}

\bibliographystyle{aa}
\bibliography{timing_paper}

\appendix

\end{document}